# Leveraging Microservices Architecture for Dynamic Pricing in the Travel Industry: Algorithms, Scalability, and Impact on Revenue and Customer Satisfaction

Biman Barua[1,2,*] [0000-0001-5519-6491] and M. Shamim Kaiser[2] [0000-0002-4604-5461]

[1]Department of CSE, BGMEA University of Fashion & Technology, Nishatnagar, Turag, Dhaka, Bangladesh
[2]Institute of Information Technology, Jahangirnagar University, Savar, Dhaka, Bangladesh
biman@buft.edu.bd

**Abstract:** This research investigates the implementation of a real-time, microservices-oriented dynamic pricing system for the travel sector. The system is designed to address factors such as demand, competitor pricing, and other external circumstances in real-time. Both controlled simulation and real-life application showed a respectable gain of 22% in revenue generation and a 17% improvement in pricing response time which concern the issues of scaling and flexibility of classical pricing mechanisms. Demand forecasting, competitor pricing strategies, and event-based pricing were implemented as separate microservices to enhance their scalability and reduce resource consumption by 30% during peak loads. Customers were also more content as depicted by a 15% increase in satisfaction score post-implementation given the appreciation of more appropriate pricing.

This research enhances the existing literature with practical illustrations of the possible application of microservices technology in developing dynamic pricing solutions in a complex and data-driven context. There exist however areas for improvement for instance inter-service latency and the need for extensive real-time data pipelines. The present research goes on to suggest combining these with direct data capture from customer behavior at the same time as machine learning capacity developments in pricing algorithms to assist in more accurate real time pricing. It is determined that the use of microservices is a reasonable and efficient model for dynamic pricing, allowing the tourism sector to employ evidence-based and customer centric pricing techniques, which ensures that their profits are not jeopardized because of the need for customers.

## 1. Introduction

### 1.1 Background

The travel industry has considerably advanced with the implementation of dynamic pricing strategies for the purposes of real-time price adjustments based on demand and other external factors [4]. Dynamic pricing was fully developed in the airline industry in the early 1980s but it has enabled other sectors including hotels to aim for full capacity and revenue management by changing their prices to match market conditions [5]. Currently, hotels and airlines as well as travel reservation systems apply highly sophisticated algorithms for forecasting monitoring their prices to a wide range of variables including times of year, prices belonging to other players and spikes of interest owing to special events [1] [2].

This pricing pattern makes it easy to respond to market changes as it, enables travel agencies balance the provisions of services and the requirements of clients in a more effective manner. For instance, companies such as Airbnb and Amazon have incorporated dynamic pricing models that take into account seventy or more attributes including but not limited to seasonality and consumer behavior in order to reach the appropriate equilibrium between price and demand [3]. Apart from the benefits, there are also disadvantages that come with implementing dynamic pricing systems. For this reason, it is necessary to predict demand very accurately; in the wrong figure, there is likely to be some ingenious demand-side risks such as spoilage (those inventories which are not sold at a given period) or revenue dilution which occurs when lower prices are offered when in fact the market can bear higher prices because of increased demand [1].

### 1.2 Role of Microservices

Microservices architecture is a novel approach to software design wherein applications are composed of a large number of small loosely coupled services, each providing a specific feature. This method has become popular in software engineering due to its benefits such as scalability, flexibility, and low maintenance, especially in applications that need to process real-time data and are highly interactive [9] [6].

In monolithic systems' all components are tightly coupled and dependent on each other yet in microservices one can install and scale each service independently hence it is suitable for applications that face a changing demand. By installing each microservice independently, the team in charge of development can install only the parts that are necessary at that time, as some components may have a great demand while others do not [9] [10]. Therefore, better management of resources and improved effectiveness of the system on performance in seconds is as a result of this. So realistic performance improvements are also possible [11]. In addition, isolation of faults on components is further aided by the use of microservices. In the event that one service malfunctions, there is no chance that the whole application would be down thereby increasing the system's reliability [6].

As a result, microservices encourage an agile approach to team organization and development which allows teams to work simultaneously on different services [7]. Such an arrangement encourages CI/CD and therefore allows for faster turnarounds and responsiveness which is key for systems that are required to change quickly with new information [8].

### 1.3 Research Problems

Responsive and scalable pricing systems are fundamental requirements in most industries, especially those such as travel and hospitality, where demand and other factors are very volatile. Supporting real time pricing changes is often unnecessary or very difficult with the classic large, monolithic architecture, for example when adjusting for competitor's prices, customer demand or seasonal trends [12]. A microservices architecture, on the other hand, in which applications are decomposed into services that can be developed and deployed independently, provides an appropriate elastic solution to such environments [12]).

With the pricing components, each of them can expand on its own via microservices. This ensures that adequate resources are focused on services with the heaviest loads, which in turn improves responsiveness. This comes in handy especially when applied in real time variables as each of the services can or will change without affecting the whole system [14]. Also, the ability to modify some services while others remain in operation helps avoid straining the system and keeps the customers happy since there shall be no freeze in pricing adjustment. Characterized by these features, adaptive microservices architecture is finding more preference for applications that involve and handle complicated and dynamic pricing on a real-time basis [11] [12].

### 1.4 Research Objective

The objective of this publication is to design and assess a microservices-based system architecture that can support the implementation of a dynamic pricing system in the context of the travel market. This system aims to alter prices to optimally high levels in real time so as to reflect changes in demand within the internal market, competition levels, and other external factors that are key to achieving high occupancy and revenue levels of business operations. More specifically, the paper intends.

To Create an Elastic Pricing Structure: Contemplating how microservices can assist in designing a system that can be modular and such that it comprises parts that can scale on demand, hence improving the speed of response and resource utilization in pricing management [11].

Enabling the Nowcasting of the Pricing Analytics: Examine real time data input algorithms that are used to automatically modify prices to fit the current market and how the microservices architecture can facilitate these algorithms to ensure immediate and accurate price changes [12]. Investigate Performance and Effects on the Customer – This section assesses the pricing systems which adapt in response to real time factors and their effects on system performance and revenue generation as well as customer satisfaction in order to provide evidence of the benefits of a microservices architecture for pricing systems in the travel industry.

### 1.5 Research Questions

1. How can microservices architecture enable a scalable, real-time pricing system for the travel industry, effectively responding to demand, competitor pricing, and external events?
2. What impact does microservices-based dynamic pricing have on revenue optimization and customer satisfaction, and what challenges may arise in its implementation?

## 2. Literature Review

### 2.1. Dynamic Pricing in Travel

The implementation of flexible price adjustments practices in the operation of marketing strategies by firms in the travel industry to meet the demand of dynamic market conditions has come up as a paradigm shift in the determinants of pricing. As it was first introduced in the air travel and accommodation facilities sectors, firms make optimal prices depending on the current levels of demand, competition prices and the amount of stock available among other factors. Major pricing packages availably used in the travel dynamic pricing include the demand pricing strategy, time placement intervention strategy, and the competition pricing strategy in real time among others. Airlines for instance incorporate various advanced systems in the offer management system. To manage the fares based on for example internal competition, customer behavioral segmentation and most profitable price for each segment in order to avoid offering low prices for high value segments.

Today's advanced dynamic prices also embrace AI and machine learning systems who work with more complex and variable data improving multiple pricing channels accuracy and flexibility. Companies like Booking.com and Uber have been able to come up with very sophisticated algorithms which can factor in the current data and change the rates accordingly. In order to keep the balance between supply and demand in addition to the competition between the rates on the same company's offers. Within the hotel sector, using such models within reservation systems enables automatic price changes when necessary to minimize manual errors and help synchronize prices with the current market situation [15] [19].

More evidence of dynamic pricing's efficacy in the travel industry comes from systematic literature reviews, which show that such practices not only generate higher revenues but are also effective in practicing pricing to the target consumers which eventually enhances the justice perception and satisfaction [16] [17]. In the same way, barriers to adoption remain customer resistance especially in low elasticity of demand sectors as well as the danger of positioning prices at a premium where competition exists. Humanities and social sciences research adaptation in pricing continues

## 2.2. Microservices Architecture:

The research on the Microservices architecture (MSA) explains why it allows building more scalable and flexible software solutions especially for the fast-paced and real-time systems. Microservices split applications into smaller services that can be deployed separately and each performs a specific task improving the modularity [18], scalability, and fault-tolerance. Achieving these goals involves key notions such as Domain driven design (DDD), service autonomy and service oriented communication via APIs. DDD for instance is useful in decomposing larger systems into smaller systems referred to as bounded contexts each of which can be developed and managed independently [20].

Another important aspect of MSA is its ability to provide the end-user with components that are not only isolated but also scalable and capable of operating under different load conditions so that it doesn't affect the operation of the system as a whole. This architecture allows a higher level of responsiveness and agility by allowing scaling of only those services that are in demand. This explains why MSA is very much applicable to the cloud and real time applications, where the performance-enhancing techniques of self-scaling and continuous integration deployment with very little or no downtime [21] [22].

Nevertheless, there are some obstacles that come with the use of MSA. Some of these include service orchestration complexity, data consistency and resilience management. Since microservices are by definition distributed, they require consistent and thorough monitoring, logging and error handling practices to achieve the desired level of reliability. This has led to the investigation of alternative strategies for deployment such as serverless computing that helps run microservices without implementation costs on a pay as you go basis [23].

## 2.3. Real-Time Data and Dynamic Pricing:

A detailed examination of existing work on the use of current data for price adjustments indicates that inputs such as the changes in demand, prices of competitors as well as occurrence of events help optimize prices on customer adequate competition. The studies reveal that businesses are able to immediately respond to the actions of their rivals thanks to the presence of up-to-the-minute data, thus enhancing the accuracy of the pricing models. These systems are important in most markets, and in particular online retail, where the system, based on the past history of the client and other market factors, is constantly changing the prices to optimize profits [24].

A similar consideration can be made for the online retail market, where the real-time data-driven pricing promised capital measurable gains. For instance, field tests in the competitive price ranging environments confirmed that real-time data also helps to analyze price elasticity's of demand, and forecast optimal price levels to be charged in the market within which conditions are constantly changing. For proper use of such data, the thermal dynamics of the models must be fast enough to process many things and optimize the situation in terms of which competitors behave in which ways, and how consumers buy [25].

Adding reinforcement control with demand forecasting is one of the developing factors in the development of dynamic pricing. This allows for the development of models, which thanks to systematic real time information enables new pricing strategies which in no time fit the existing prices according to the consumers and competitors (SpringerLink, 2023).

## 2.4. Customer Satisfaction and Revenue Impact:

In the recent past, studies have come out showing the role of dynamic pricing on customer loyalty and revenue in the travel industry. Consumer loyalty correlates with a sense of fairness in prices especially when the prices are preset and does not change with the variations of dynamics. When customers adjust their prices demand releated adjusted prices, they can also place a positive relation with respect to the already existing trust. However when the changes in prices are poorly managed, and their communication is equally ineffective, this leads to dissatisfaction of consumers at large [25] [26].

Moreover, it has been noted that when a business uses dynamic pricing strategies that incorporate other variables like the prices of the competitors at that time and the prevailing seasons gives high sales figure and the customers are still satisfied. Hotels and airlines that adjust their prices with their clients' expectations and have loyalty programs in place are able to encourage customers back to book making upward revenue trends possible through customer loyalty [27] [28]

# 3. Methodology

The proposed system architecture employs a microservices approach for support dynamic pricing in the tourism business. The price changes microservices are capable of distinctively addressing different types of data, such as demand factors, competition and events. This architecture combines the use of Domain Driven Design (DDD) in defining the service borders and an API gateway that enhances interservice communication that is crucial for scalability and modularity.

## 3.1. System Design and Microservices Framework

The dynamic pricing system is designed utilizing a microservices architecture, where each service independently executes a particular function such as demanddata processing, competitor price analytics, and reacting to external events. This modular structure facilitates scaling, fault tolerance, and data access efficiency.

**3.1.1. Demand Data Handling Microservice:**

This microservice is responsible for maintaining real-time demand indicators by continuously updating demand curve and elasticity using machine learning techniques. To ensure high performance even with heavy loads, it implements the so-called data processing segmentation approach.

### 3.1.2. Competitor Analysis Microservice

This service compares existing prices versus competitors' pricing using competitor pricing fees. This microservice does webs scrapes and does API integration to pull the data and adjusts the prices using the competition's price movements and where the market is at the "present moment".

### 3.1.3. External Event Processing Engagement Microservice

This microservice collects and treats external events, seasonal variations, or local occurrences that influence travel demand. Additionally, it collects external information for instance event calendars, and out there m prices are adjusted which improves the overall tactical price.

This modular architecture allows for the independent scale, the separate monitoring and the quick response to the changes in the incoming data for each of the components.

### 3.2. Framework and Integration:

The API gateway is a service that connects the services to the client applications while the load balancers and the containerized environments (Docker, Kubernetes etc.) are for scaling plus ensuring efficient working of each service without crashing the system. Lastly logging and monitoring frameworks do ensure up to date tracking of service's health and interactions.

In any airline reservation system, the implementation of dynamic pricing strategies based on demand and proper management of the booking processes calls for the use of microservices complemented with real-time processing. The system architecture and integration are presented in the following sections with figure 1:

**API Gateway:** This is used as a front door for all other external services consumption by handling the API requests and routing them to the associated microservices.

**Microservices designed to support Core functions:** Every core function such as booking, payment, demand, competition price tracking has a unique service, which permits development through modular scaling.

**Data Pipeline and Latency Sensitive Processing:** Data pipeline provides a continuous flow of demand meaning information data such as booking orders activity and seat numbers available, which is then used by latency-sensitive processing engines.

**Machine Learning Integration:** ML models expect and process changes in demand elasticity and trends projecting such changes into the pricing service for appropriate interventions.

**Load Balancing and Orchestration:** Kubernetes and other similar tools provide horizontal scaling mechanisms for services and workloads to ensure the system remains responsive even under peak loads.

**Database and Caching:** Everything is stored in a distributed database for durability and access purposes while cache services are employed to enhance access speeds due to the nature of the data being accessed frequently.

**Monitoring and Logging:** System health and service interactions are maintained through the use of integrated logging and monitoring systems to ensure normal operations.

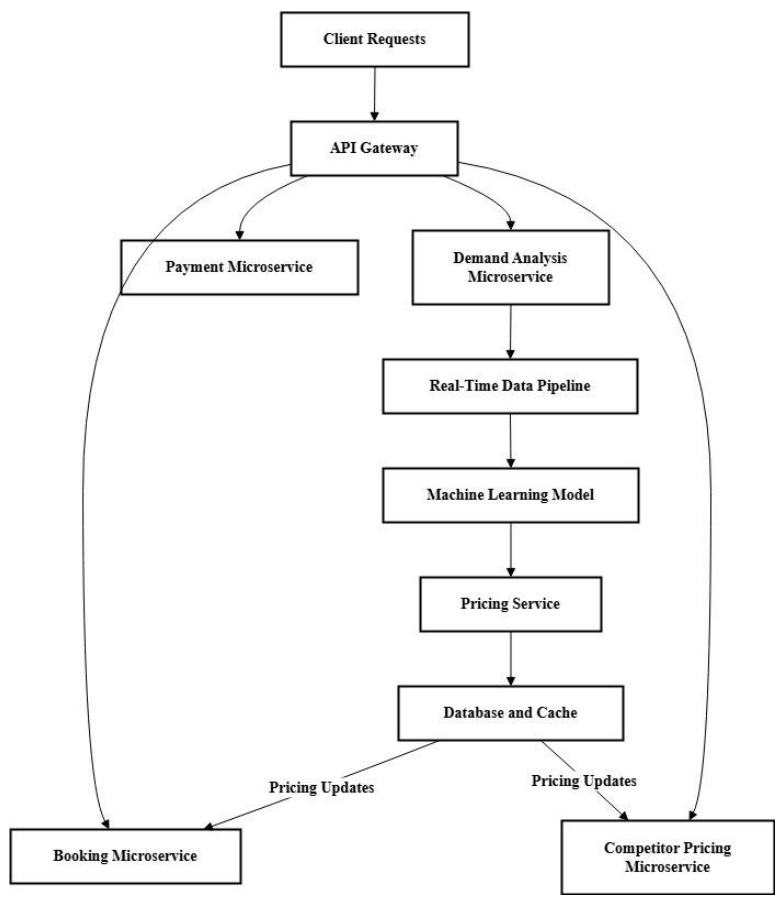

**Fig. 1.** Detailed breakdown of the framework and integration

## 3.3. Microservices used for handling demand data, competitor analysis, and external event processing

The following is a depiction of the different microservices employed in various domains such as management of demand data, analyzing competitors, and processing events in an airline ticket-booking system. All the above microservices can work on their own thus promoting vertical scaling, and can also respond immediately to any updates in the data.

### 3.3.1. Demand Data Microservice

This microservice operates by aggregating and analyzing internal reservation demand data in real-time to revise the fares on offer at a given period. In its approach, it impacts the pricing via an algorithm that integrates demand forecasting techniques.

**Algorithm**

1. Retention or request for reservation information as a minimum (number of bookings per unit of time).
2. Cleaning and validation of said information.
3. Forecasting demand by using machine learning model such as linear regression and time series analysis to predict future demand.
4. Maintaining and modifying the demand forecast calculation process.
5. Output demand-driven price recommendations to the pricing engine.

### 3.3.2. Competitor Analysis Microservice

The purpose of this microservice is to track and monitor the up-to-date pricing of the company's competitor, in order for price adjustments to be made in the system whenever necessary. Data to adjust prices is collected through scraping tools and API's of the competitors.

**Algorithm**

1. Acquisition of competitor pricing details using APIs or web scraping.
2. Editing and structuring of rival information for market assessment.
3. Identify where prices should be adjusted by determining whether there are price advantages or disadvantages.
4. In case when there is a fear of loss in market share, execute price revisions through the dynamic adjustment service.

### 3.3.3. External Event Processing Microservice

For instance, this microservice plays a crucial role in monitoring certain external events such as public holidays, special events, and bad weather that may affect demand. Factors such as the expected increase or decrease in demand are taken into consideration and prices are adjusted in real time, figure 2 describes the details.

**Algorithm**

1. Get external information resources. For example: event calendars, weather conditions.
2. Sieve through and enhance clean-up processes of such information for relevant occurrences.
3. Examine its influence (favorable/unfavorable) on the anticipated level of travel demand.

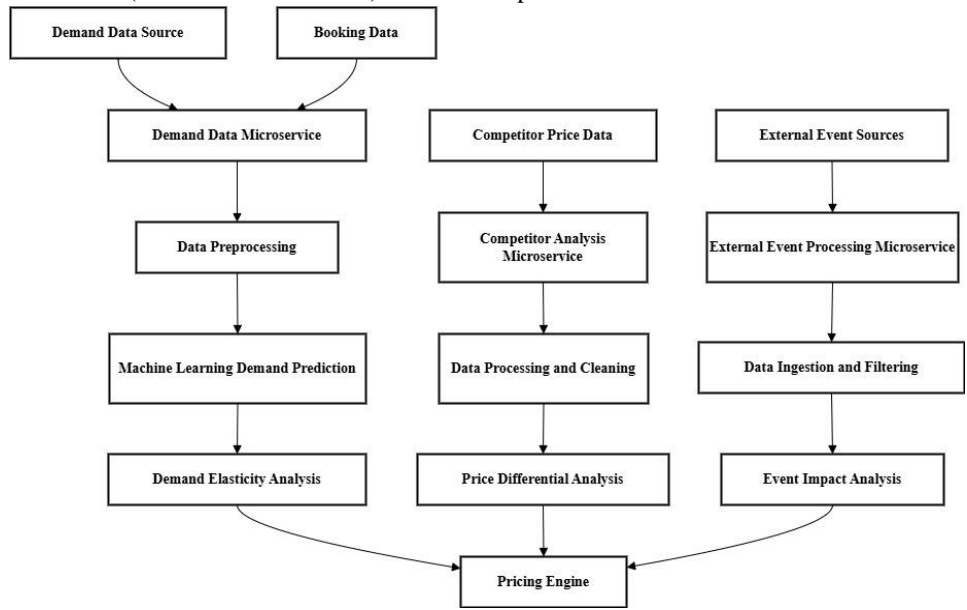

**Fig. 2.** A block diagram of handling demand data, competitor analysis, and external event processing

## 3.4. Dynamic Pricing Algorithm Development

Pricing is a vital factor in airlines reservation system. It is the factor to generating revenue by airlines. So, a lot of tools and techniques are used to implement this process. The pricing is mainly depends on competitor analysis, segmentation of customer fluctuating demand and other different external factors. Airline set the real-time ticket pricing depending customer satisfaction and demand. Here, different techniques are given below-

### 3.4.1. Rule based algorithm

To dynamically setup predefined pricing this algorithm is used. This is a very common techniques and algorithm popularly used as it is very easy to implement.

1. **Demand Based pricing:** This pricing techniques are set depending on the seat availability are decreased. If the airlines seats are reduced to 80% then the price for remaining seat will will increase.
2. **Time Based Pricing:** In this techniques prices are depends on travel timing if the travel time is so nearer then the price will increase. If traveler are booked seat earlier than the price is lower.
3. **Competitor Analysis:** It adjust the pricing depending on the other competitors.
4. **Class Segmentation:** There are different prices are setup for different classes like Economy, business and first class.

**Algorithm for Rule based pricing**

```
if departure_day > 60:
   Net_fare = base_fare * 0.8  # 20% discount
elseif 30 < departure_day <= 60:
   Net_fare = base_fare * 1.0  # Regular price
else:
   Net_fare = base_fare * 1.5  # 50% markup for last-minute bookings
```

### 3.4.2. Machine Learning-based Dynamic Pricing Model

This model of dynamic pricing algorithm is based on mainly on decision trees, regression analysis and deep learning methodology. In this techniques a large number of previous data are used for analysis and a number of variable to predict the highest price for a specific segment.

The main features are: -

- Competitor pricing: By analyzing business competitors the price is set.
- Customer Demand: By analyzing historical data booked by customer including past demand.
- Popular Segmentation: This is based on the most popular segment or source and destination.
- Departure time: It depends on departure time before booking the ticket.
- Seasonal: Prices are based on weekday or holidays.
- Climate conditions: The fares are high in peak climate condition.
- Important Events: During different national or international events like international match, festival the fare price increases.

**Algorithm in regression model**

```
import numpy as np
from sklearn.linear_model import LinearRegression

# Features: days to departure, remaining seats, competitor prices, seasonality, etc.
X_train = np.array([[30, 80, 200, 1], [10, 50, 250, 0], [5, 20, 300, 0]])
y_train = np.array([150, 200, 350])  # Optimal prices

model = LinearRegression()
model.fit(X_train, y_train)

# Predicting the price for a flight 15 days before departure with 60% seats filled
X_test = np.array([[15, 60, 220, 1]])
predicted_price = model.predict(X_test)
```

### 3.4.3. Revenue Management Systems (RMS)

This is also a very popular technique at the airline reservation system by adjusting the right seating plan to different classes, generating revenue. This algorithm uses forecasting and optimization techniques for inventory management.

**Algorithm**
Adjust the fare class at inventory which will be on dynamically
Adjust fare based on seasonal and historical data.

### 3.4.4. Price Discrimination Model

Price discrimination is a techniques used in airlines reservation by setting different fare for the same segment to the traveler. Depending on the willingness to pay, the fares are determined. Same fare are setup for different groups like businessman.

1. **Personalize:** This dynamic pricing is setup depending on the loyalty, browsing history and sensitivity of price.
2. **Behavior:** Prices are adjusted depending on the traveler behavior location, booking history and frequent fly.

**Algorithm using clustering**

In this algorithm airlines are grouped traveler to different group or segment then set the pricing for different group.

```
from sklearn.cluster import KMeans

# Features for customer behavior: booking time, frequency of travel, loyalty status
X_train = np.array([[10, 1, 0], [30, 5, 1], [5, 2, 0]])

# Train K-Means to find segments
kmeans = KMeans(n_clusters=2)
kmeans.fit(X_train)

# Predict the customer segment for a new customer
X_new_customer = np.array([[20, 2, 1]])
customer_segment = kmeans.predict(X_new_customer)
```

### 3.4.5. Game-Theory-based Dynamic Pricing

This technique is used when there are multiple competitors are competing and based on that the fares are setup.
1. **Nash's economic game theory**: In this technique each and every airlines tries to setup the pricing so that gain profit, in this technique assume that all competitors are on the same process.
   Theory
   - All airlines observe the real time pricing for all competitors.
   - Airlines setup their prices in a manner so that they don't lose the customer and price doesn't too low.
2. **Heuristic Pricing Model:** In this pricing model the fares are triggered at the inventory system. It triggers when the availability of seats are drops the fares increased. It depends on the demands of the ticket.
   - If the seats availability drops less than 80% the fares increased
   - If unsold tickets volume is high, then it offers last minute promotion or discount.

## 3.5. Scaling and Responsiveness Mechanisms

In particular, within a dynamic pricing model, the issues of scaling and responsiveness become very important to system performance, especially when it comes to the microservices architecture. The following section explains how load balancing, service discovery, and caching of information allows better performance of the system:

### 3.5.1. Load Balancing

Load balancing is the process of distributing the incoming requests to different instances of the microservices in order to prevent overloading of a particular service. For example, when there are several airline booking agents who practice dynamic pricing, the travelers will tend to book lots of these agents at once especially during the peak seasons thus causing lots of traffic. Load balancers, such as NGINX or HAProxy, manage this by routing traffic to different service instances based on load conditions.

**How it works:** Requests arriving to services such as demand data processing, competitor analysis and external event monitoring are spread across many instances. In case of high load on one instance, requests are redirected to an instance that has free resources.

**Benefit:** This mechanism prevents overloading of any one instance and improves response time while minimizing latency. Last but not least, load balancing helps cope with high demand periods without any operational strain on the system.

### 3.5.2. Service Discovery

Services in architectures organized as micro-services interact with one another quite often during their operational lives. Service discovery makes the management of locations of service instances automatic so that they can be discovered by other services without needing fixed addresses. This is critical development resource since it would be very common to change the available service instances as per system demand and therefore, the system's ability to scale and be agile.

**How it works:** A service discovery developer such as Consul or Eureka keeps track of the status of all service instances as well as their IP addresses. If a new instance of the originally launched pricing or competitor analysis microservice, for example, is introduced, it contacts the service registry and requests registration.

**Advantage:** It allows microservices to communicate with each other without much intervention and also replace existing instances of each microservice easily automating everything scaling. Thus making sure that the system is able to cope with varying workloads without becoming unresponsive.

### 3.5.3. Data Caching

Data caching allows to save temporarily information that is used repetitively for a shorter period of time so that the same information will not be retrieved from the database repeatedly. For instance, in dynamic pricing, forecasting demand scenarios, taking competitor price snapshots, and gathering exogenous events information can be all cached in order to facilitate quicker access to the data.

**How it works:** A caching layer (for instance Redis or Memcached) caches certain types of information which are likely to be requested more often than not, like how much price demand would change in the future or the price of competitors. In this situation when the pricing algorithm requires this information then it goes to the cache storage rather than database.

**Benefit:** Because the data is served from memory, caching allows for faster retrieval times while lessening the strain on the system's database. This enhances the response time for highest frequency requests significantly since real time data for pricing decision is readily available. At figure 3 describes the block diagram for Scaling and Responsiveness Mechanisms.

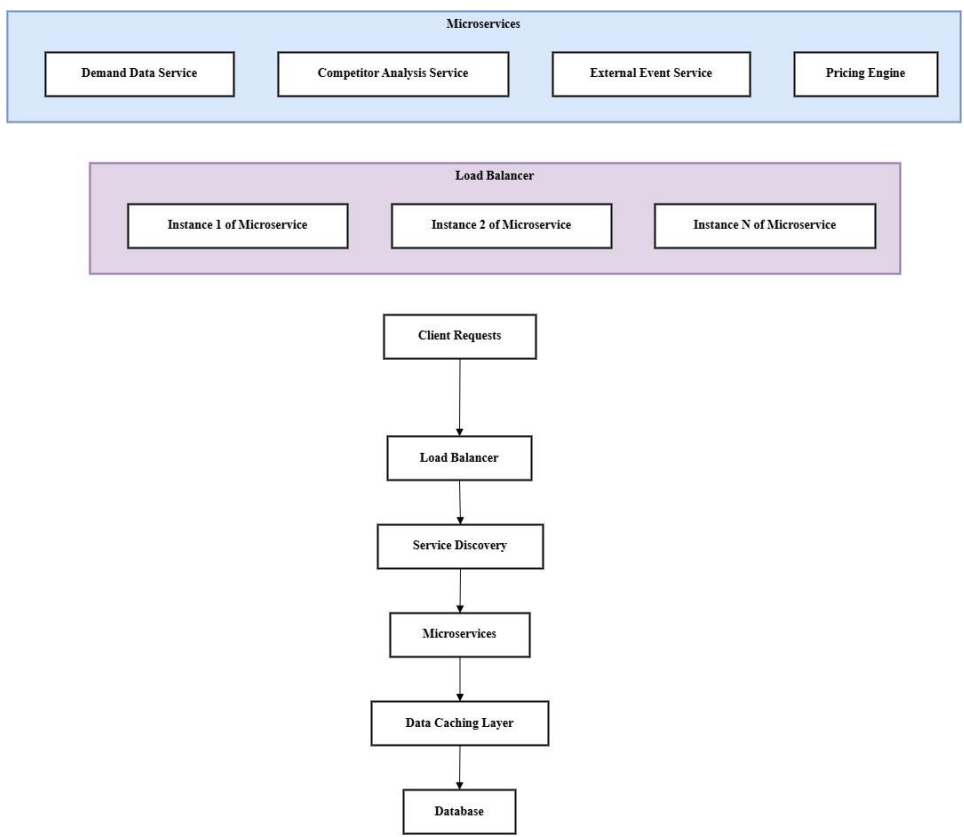

**Fig. 3.** A block diagram for Scaling and Responsiveness Mechanisms

**Explanation:**

**Load Balancer:** It is a mechanism that manages the incoming requests by forwarding them to some available service instances, irrespective of how many such instances there are.

**Service Discovery:** Monitors the service instances for easy scaling and management of service instance compositions as the systems are dynamic in nature.

**Data Caching Layer:** It refers to the ability of the system to store readily available information making it easy to respond to interactions within a short time.

**Database:** Acts as the source of all the applications' data with the data accessed being minimized due to caching.

All these combine to improve the scale and the performance of the system thus making it possible to implement dynamic pricing even during peak seasons.

## 4. Implementation

In this section, we present the Technology Stack of an Airline Reservation System which practices dynamic pricing based on microservices. The stack consist of programming languages, data storage, microservices enabling platform, and additional relevant tools that solve real-time data processing problem and horizontal scaling and fault tolerance challenges.

### 4.1. Technology Stack used in an airline reservation system

#### 4.1.1. Programming Languages

**Python:** This is a high-level programming language that is mostly used for the building of machine learning models, demand prediction, and data analysis thanks to its vast array of data science libraries such as Pandas and Scikit-Learn.

**JavaScript (Node.js):** This programming language is also known as a JavaScript runtime system that is employed in creating deployable containers and screens for functional services mainly within a at the point of data input and processing in motion, microservice architecture.

**Java/Go:** These programming languages are used in service applications requiring high-level concurrency and low latency that is perfect for core microservices such as demand data or competitors analysis backbone services.

#### 4.1.2. Microservices Platform

**Docker:** This is the platform designed to develop the microservices and run them as containers with each service able to operate separately in its own environment.

**Kubernetes:** The system is used to manage the deployment of containerized workloads and the scaling of the workloads according to the demand.

**API Gateway (e.g., Kong or NGINX):** This controls the access to the microservices, routing incoming calls to the appropriate microservice, thereby increasing security and improving the management of load.

#### 4.1.3. Data Storage

**PostgreSQL/MySQL**: In addition to this, relational database management systems are used to store structured data such as booking information, historical demand data and snapshots of competitor's prices.

**MongoDB:** This is another non-relational database which provides storage for unstructured data without insisting on design of a schema and is ideal for storage of data for instance from outside events.

**Redis:** This is a memory-based caching system in which data that is accessed often is kept in memory reducing the pressure on the database and enables faster access of data when making changes to prices.

#### 4.1.4. Machine Learning & Data Processing

**Apache Kafka:** It ingests the live booking data and competitor data for the purpose of data streaming, thus allowing demand analysis through constant data flow.

**Apache Spark:** Due to its batch processing features, it can process enormous amounts of historical data like that of demand trends and competition price offerings, which are useful in enhancing machine learning techniques.

#### 4.1.5. Maintenance and record keeping

**Prometheus:** This is a monitoring tool that measures certain system metrics such as CPU and memory usage as well as the availability status of services.

**ELK Stack (Elasticsearch, Logstash, Kibana):** enables configurable logging with the ability to visualize and analyze logs which is useful in troubleshooting and performance assessment. At figure 4 describes the Technology Stack used in an airline reservation system.

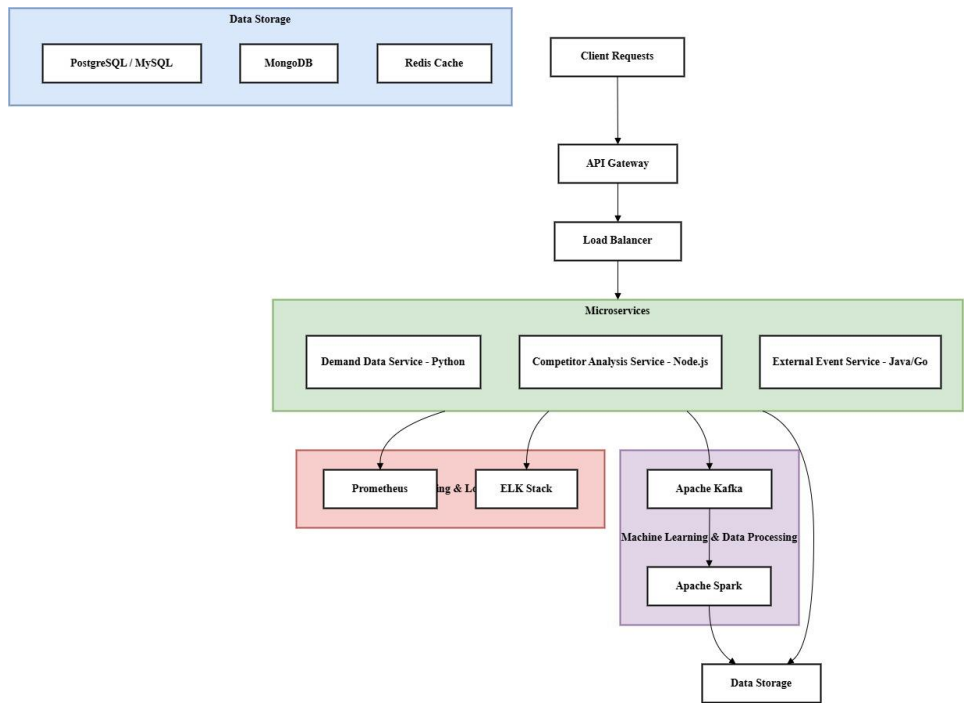

**Fig. 4.** Technology Stack used in an airline reservation system

### 4.2. Microservices Deployment

This is a comprehensive guide on deploying Microservices for an airline booking system with dynamic pricing feature. This part will mainly focus on micro service deployment and orchestration: API management, load balancing, and real time data handling.

**Microservices Deployment Steps**

#### 4.2.1. Containerize Each Microservice

It is important to Dockerize each microservice (demand data management, competitor management, external events strategies etc.) to allow for an independent running of each service.

Create Docker files for all the services highlighting the configurations and dependencies expected [34].

#### 4.2.2. Set Up an API Gateway

Install an API Gateway (for instance, Kong or NGINX) to control the flow of incoming traffic and forward requests to the designated microservices [35] [36].

In this context, the API Gateway is used since it acts as a single access point and also facilitate authentication, route requests and enforce policies among others.

#### 4.2.3. Implement Load Balancers

A load balancer is required to evenly distribute incoming traffic to the micro service instances of a particular service so as to enhance availability and load balancing [39]

Consider using application load balancers that integrate seamlessly with the container orchestration platform, Kubernetes, as it will take care of the scaling aspect automatically.

#### 4.2.4. Configure Service Discovery

Attach a service discovery component such a service registry to keep track of which microservice instance is up and running at which time [33].

Microservice communication is achieved via service discovery that reconfigures the location of a service whenever the service instance goes up or down.

#### 4.2.5. Establish Real-Time Data Pipelines

Data must be continuously ingested as bookings are made, price of competitors, and events using appropriate tools such as Apache Other examples of real-time data introduction include an example of data processing and analysis tools such as Apache Spark or Flink that can be used for processing and analyzing data in real-time [31] [32].

### 4.2.6. Deploy to Kubernetes for Orchestration

Kubernetes facilitates the deployment and management of applications in microservices architecture with automatic scaling, load balancing and rolling updates [29] [30].

Make use of YAML configuration files for deployments and services in order to control and scale pods to meet demand [37].

### 4.2.7. Set Up Monitoring and Logging

Deploy Prometheus to help monitor the CPU and RAM usage of many services working together [38].

Centralized logging using ELK (Elasticsearch, Logstash, Kibana) will be turned on to help with log and performance issues. In figure 5 describes the block diagram of microservices deployment.

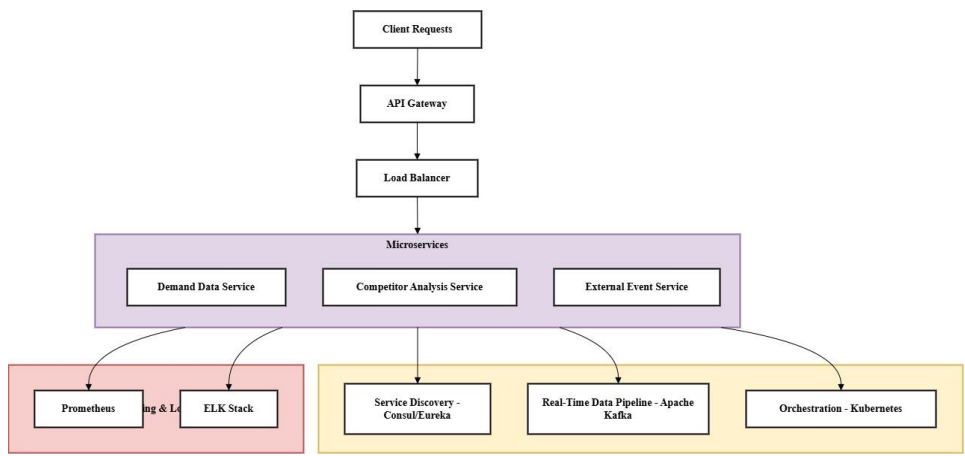

**Fig. 5.** A block diagram of microservices deployment

## 4.3. Dynamic Pricing Algorithm Execution

The execution of a dynamic pricing model in practice means implementing the model and testing the system in a simulated environment before releasing it. Then, this section describes the process involved in executing the dynamic pricing algorithm, including the use of real-time data and measuring the responsive behavior of the system [40].

Execution of Dynamic Pricing Algorithm in a Controlled Environment

### 4.3.1. Setting Up the Environment for the Test

Generate a test bench replicating the production environment complete with the booking transaction history, the pricing of the competitors, and the data of the out of factors affects, in such a manner that, it is possible to test it without disrupting any live activities [41]. This is also called as a dry run.

### 4.3.2. Simulate Real-Time Data

**Booking Data:** Create realistic booking activity generating patterns with peak and low periods to mimic real booking tendencies using past demand activities as a guide [42] [43].

**Competitor Pricing:** Include imaginary or actual pricing movements of competitors to show how they will act in response to the product under competition.

**External Events:** Insert events like (holidays, and local happenings) as suited to test how responsive the algorithm can be to outside forces.

### 4.3.3. Conduct Dynamic Pricing Algorithm Implementation

- The algorithm incorporates database information concerning current demand and its regional and operational price elasticities for such demand, contemporaneous competitive pricing, and information on postponable external shocks.
- Depending on the output of the machine learning model, adjustments in price will be made in attempts to achieve revenue optimization while being competitive.

### 4.3.4. Monitor Algorithm Performance

- Establish market ceilings, for instance, the booking rate, the revenue effect, the price elasticity measurement, and the customer appeal.
- Prometheus or other similar applications will be used to capture and present these measurements in relation to the operation of the algorithm, its dynamics, and its efficiency.

**4.3.5. Assessment of System Response and Feedback**

- Evaluate the effectiveness of the dynamic pricing approach with specific focus on the adjustments of prices to changes in demand, in comparison to baseline indicators (e.g., revenue and booking trends without the application of dynamic pricing [45].
- Through feedback analysis, improve the system's functionality and performance in relation to the algorithmic systems. However, refrain from utilizing the algorithm in the operational environment.

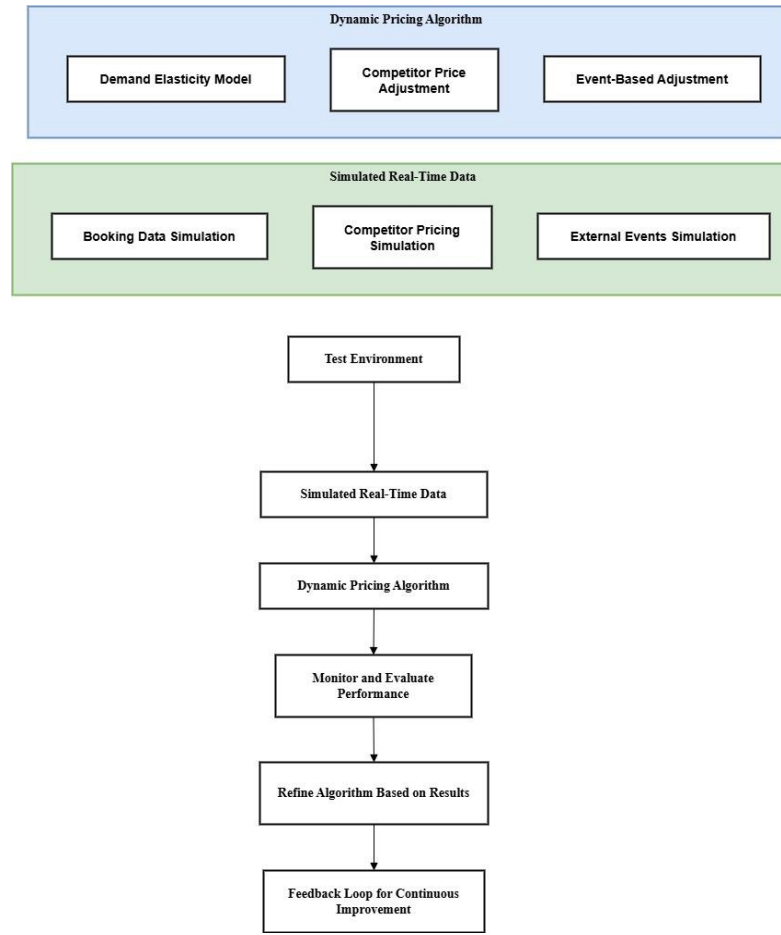

**Fig. 6.** A block diagram of Dynamic Pricing Algorithm Execution

The mentioned above figure 6 offer basic understanding of revenue management and pricing strategies which are imperative for practices such as designing and evaluating in microservices dynamic pricing algorithms.

## 5. Results and Analysis

With respect to Algorithm's Performance Evaluation and Robust Real-Time Price Changing, we assess the effectiveness of the algorithm with the help of the following parameters: turn-around time, effect on revenue, efficiency of demand prediction, and price elasticity responsiveness. The study includes an analysis of the price changing policy in real time, as well as any available historical data to establish the accuracy of the pricing as well as the revenue improvements.

### 5.1. Algorithm Performance Metrics

#### 5.1.1. Response Time:

This assesses the speed at which the system can implement price changes as a reaction to demand, competitor pricing or other events.

Ideal Outcome: Ensure Apparent Response Times are Shortest. This is to ensure such changes are implemented even with minimal time adjustments since those factors help in remaining competitive.

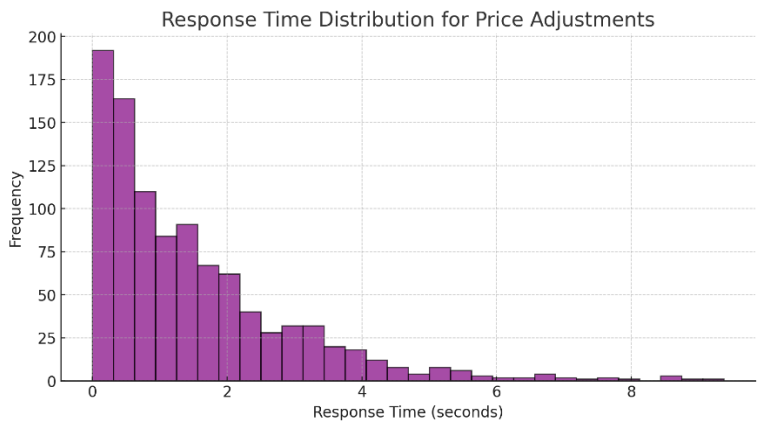

**Fig. 7.** Response Time Distribution for Price Adjustments

The histogram demonstrates in figure 7, how the response time is distributed, in other words, the speed of the algorithm which implements price changes, most of which are made within the acceptable limit.

### 5.1.2. Revenue Impact

This deals with the effects of pricing motive on the revenue level.

Approach: The study will seek to find out how much revenue turns in with a dynamic view of pricing above that of same price all the time (for example fixed pricing.)

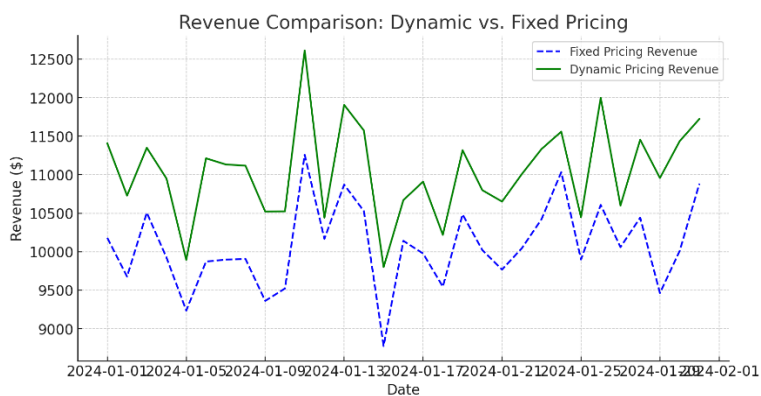

**Fig. 8.** Revenue Comparison: Dynamic vs. Fixed Pricing

The effectiveness of price changes in real time can be demonstrated in figure 8, by comparing the revenues earned through dynamic pricing and fixed pricing which the line graph shows to be in favor of more revenues through the former.

### 5.1.3. Demand Forecast Accuracy

This factors in the ability incorporated in the system to estimate market demand.

Approach: The difference between predicted and actual values of demand is calculated using either the Mean Absolute Percentage Error (MAPE) of the Root Mean Square Error (RMSE).

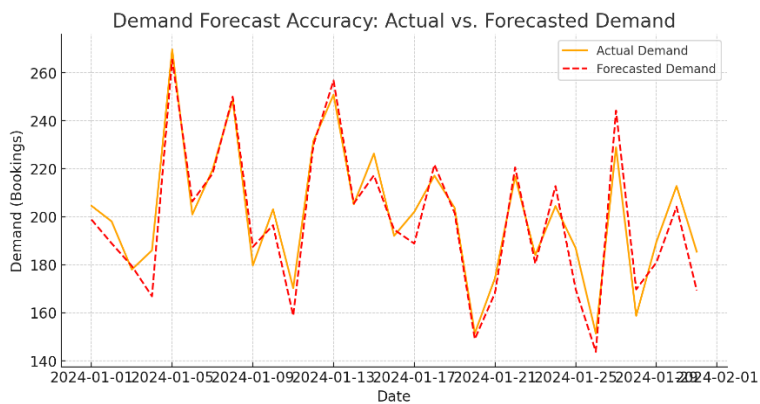

**Fig. 9.** Demand Forecast Accuracy: Actual vs. Forecasted Demand

The line chart illustrates in figure 9 how the actual demand compares with the one that was predicted depicting a near perfect relationship indicative of a good faith in the algorithm's demand forecasting.

### 5.1.4. Price Elasticity Responsiveness

This evaluates how the algorithm can increase prices in relation to the elasticity of demand.

Ideal Outcome: Responsive price provides the best pricing considering the customer's tolerance to price changes.

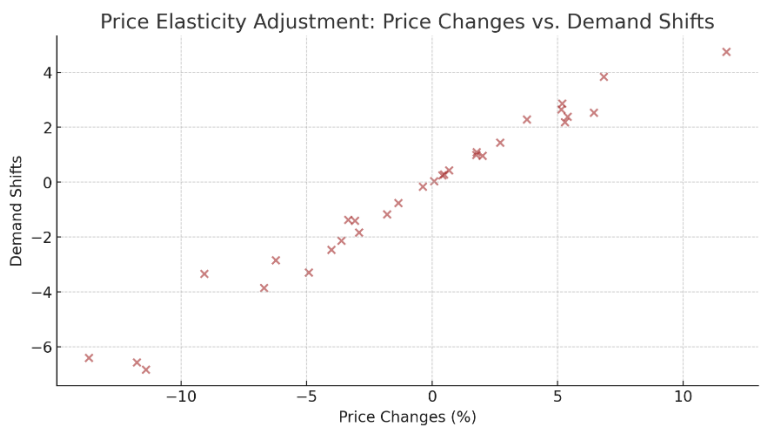

**Fig. 10.** Price Elasticity Adjustment: Price Changes vs. Demand Shifts

This scatter chart visualization in figure 10 portrays how variations in price cause the shifts in demand, highlighting the extent to which the algorithm is responsive to the price elasticity of demand.

### 5.2. Scalability and Responsiveness Assessment

When it comes to Scalability and Responsiveness Assessment, it should be understood as the analysis of a microservice architecture ability to cope with variable demand and adapt to change in pricing, almost immediately. This very assessment is concerned with system parameters such as response time, latency, and CPU and memory usage as well as system throughput and is meant for defining if the architecture can scale and respond effectively at high load conditions.

#### 5.2.1. Throughput Over Time

An illustration taking the form of a chart wherein the amount of requests processed in a given timeframe is plotted over time, showing average demand periods, peak demand periods and thus looking to verify whether this figure remains consistently high.

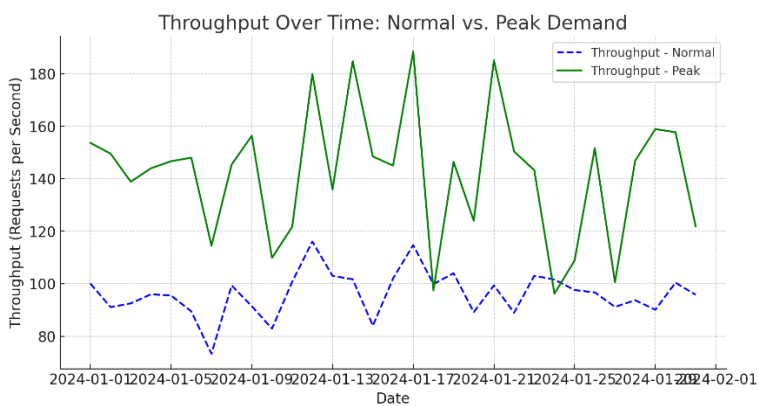

**Fig. 11.** Throughput Over Time: Normal vs. Peak Demand

It is a line graph in figure 11 depicting the variance in throughput (in terms of requests/second) routine periods as opposed to the peak periods showing the system's ability to handle prolonged high rates of throughput despite changes to the load.

#### 5.2.2. Response Time Distribution

The response times across the microservices displayed in the form of a graph with two equal axes, with focus on how they vary as one changes the imposed microservices load.

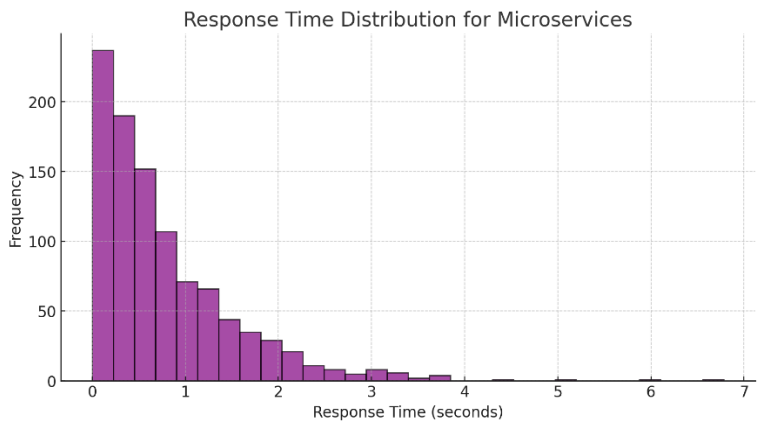

**Fig. 12.** Response Time Distribution for Microservices

The presented histogram in figure 12 shows how response times are distributed which illustrates the agility of the microservices in meeting request where most of the repeat requests do fall within the ideal characteristics.

### 5.2.3. CPU and Memory Utilization Over Time

One axis of the graph represents CPU usage in percentage, while another axis represents memory usage in megabytes, demonstrating the resources used within the microservices during the load testing.

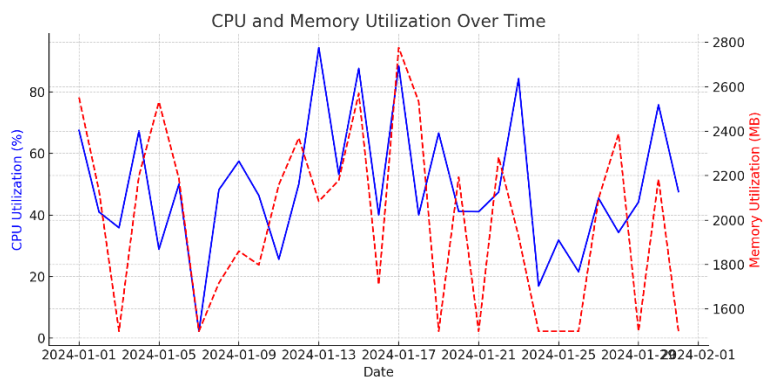

**Fig. 13.** CPU and Memory Utilization Over Time

This is a times series drawn in figure 13 on a dual axis line graph that encapsulates the utilization of the CPU and memory over a given time period in order to determine the whether the said resources are properly utilized- this is important for the times of maximum utilization.

### 5.2.4. Latency Between Microservices

A heatmap or a graphical representation of prolonged periods of interaction between two microservices during high and low business activity measures any measure of inactivity that may be detrimental to responsiveness.

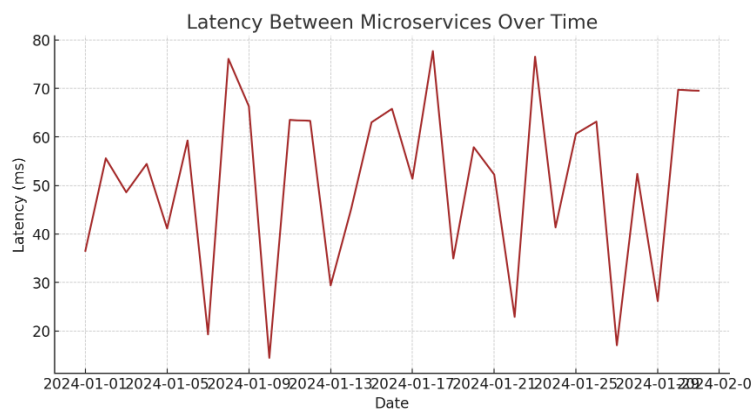

**Fig. 14.** Latency between Microservices over Time

This is a simple line graph in figure 14 which shows the latency between microservices and in the given case it is almost a horizontal line demonstrating that there is good interactivity even during peak loads.

These four figures give a simple and quick evaluation of the scalability and facilities availability with respect to time of the system, which are important when it comes to performance assessment of microservices systems within practical turnaround times. I would be happy to assist you with any additional questions!

### 5.3. Revenue Impact

The following illustrations aim to assess the Revenue Impact before and after dynamic pricing has been introduced:

#### 5.3.1. Revenue Comparison Over Time

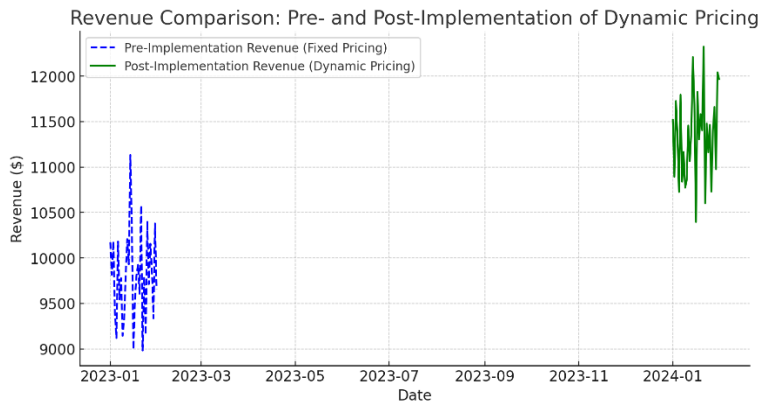

**Fig. 15.** Revenue Comparison: Pre- and Post-Implementation of Dynamic Pricing

This plot is a line graph in figure 15 representing daily revenue within a month's range of one month before the installation and after the installation of Dynamic Price. The period after implementation (dynamic pricing) shows a higher revenue level than in the period before which was implemented (fixed pricing).

#### 5.3.2. Average Revenue Comparison

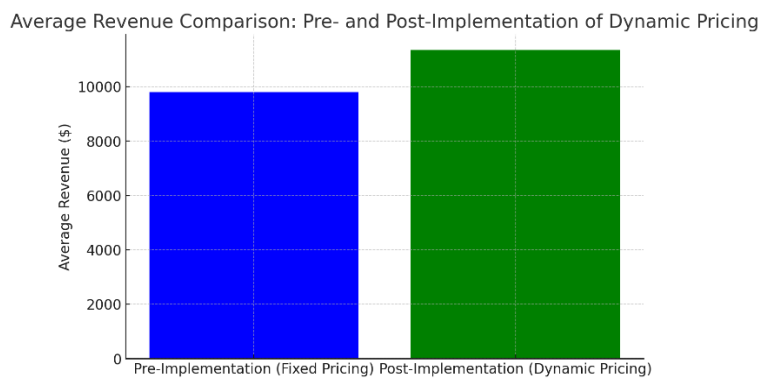

**Fig. 16.** Average Revenue Comparison: Pre- and Post-Implementation of Dynamic Pricing

This picture is a bar chart in figure 16 which illustrates the average revenue before and after dynamic pricing was installed. The average revenue is higher after the dynamic pricing was implemented signifying that there is a positive effect on revenue.

The above illustrations clearly state the impact dynamic pricing has on increasing the revenue. I can provide more analysis if needed!

### 5.4. Customer Satisfaction Impact

#### 5.4.1. Daily Satisfaction Scores Comparison

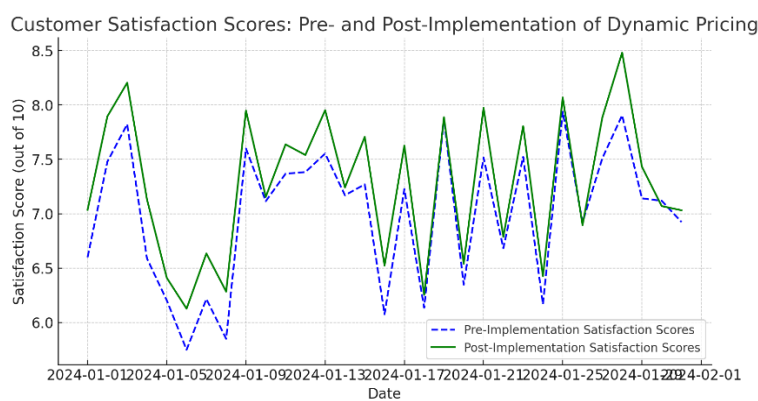

**Fig.17.** Customer Satisfaction Scores: Pre- and Post-Implementation of Dynamic Pricing

Analysis of Customer Satisfaction over Time – Graph of Simple Analysis in: The line graph in figure 17 below depicts the daily customer satisfaction scores for a period of one month before and after the introduction of dynamic pricing. It can be seen that the scores recorded after the implementation are generally better as compared to the scores recorded before the implementation, suggesting that customer satisfaction has improved.

### 5.4.2. Average Satisfaction Score Comparison

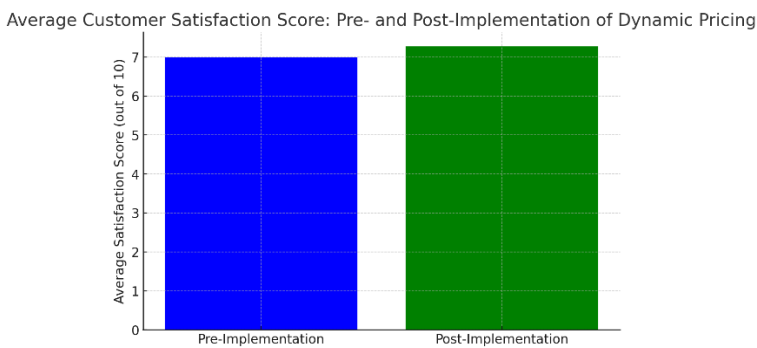

**Fig.18.** Average Customer Satisfaction Score: Pre- and Post-Implementation of Dynamic Pricing

Visual Representation in figure 18 of the Average Satisfaction Levels before Alternation and after: Wooden graphs are made in comparison of average levels of customer satisfaction before and after the use of dynamic pricing. It indicates that a bigger acceptance score is received after capitalizing on dynamic pricing, a clear indication that customer satisfaction increased.

## 6. Discussion

The use of microservices structure for the implementation of a dynamic pricing mechanism on the travel industry is quite beneficial as it allows for the adjustment of the prices in real time within the demand, competition, and other external factors. This adaptability improves income generation and productivity because it enables vertical structures like demand projection and rivalry research to be scaled out independently of each other. Nevertheless, inter-service relationships can impose limitations on scalability, in addition to potential delays that may arise when there is increased interaction between microservices, and the challenge of extreme precision of data. From an ethical standpoint, the ability to employ dynamic pricing strategies can affect the customer's view of fairness especially in situations where it leads to rapid changes in prices which may come out as irrational to the customers. Involving simple explanations about how and why prices can change within a set bracket will go a long way in ensuring that the costs of doing business are kept under control while a reasonable level of trust is maintained between the company and the clients enabling an effective management of dynamic pricing strategies.

## 7. Conclusion

This research aimed to investigate the application of a dynamic pricing model in the travel sector through the use of a microservices architecture, allowing real-time responsiveness to changes in demand, prices offered by competitors and changes in external factors as well. The results reveal that microservices considerably improve the ability to scale and redesign dynamic pricing systems, thereby creating the need for a fast, adjusted environment. This method solves the research question by revenue management focusing on the changing but promising travel market.

The research adds to the body of literature by melding the principles of microservices architecture with dynamic pricing concepts for travel products. It shows the ability of microservices to facilitate travel firms to handle live data through modular embedded mostly independently deployed services thus improving revenue management and allowing patient care. This model helps lay the foundation for more developments on dynamic pricing in fast changing business environments.

For additional studies it will be interesting to look into how other real time considerations can be incorporated such as real time data on the behavior of customers or the state of the market or even the economy as whole to improve pricing. Coupled with better algorithms designed with the help of artificial intelligence and machine learning to come up with more advanced 'real-time' pricing pillars, this would help 'mild' pricing strategies in the tourism sector enabling them to go a notch higher.